\documentclass[twocolumn,amsmath,amssymb,pre,showpacs]{revtex4}

\usepackage{graphicx}
\usepackage{dcolumn}
\usepackage{bm}

\newcommand{\bs}{\boldsymbol}

\newcommand{\mb}{\mathbf}

\begin{document}

\preprint{APS/123-QED}

\title{Alternative expression for the electromagnetic Lagrangian}

\author{Pablo L. Saldanha}\email{saldanha@fisica.ufmg.br}
\affiliation{Departamento de F\'isica, Universidade Federal de Minas Gerais, Caixa Postal 701, 30161-970, Belo Horizonte, MG, Brazil}


\date{\today}

\begin{abstract}
We reintroduce an alternative expression for the Lagrangian density that governs the interaction of a charged particle with external electromagnetic fields, proposed by Livens about one century ago. This Lagrangian is written in terms of the local superposition of the particle fields with the applied electromagnetic fields, not in terms of the particle charge and of the electromagnetic potentials as is usual. Here we show that the total Lagrangian for a set of charged particles assumes a simple elegant form with the alternative formulation, giving an aesthetic support for it. We also show that the alternative Lagrangian is equivalent to the traditional one in their domain of validity and that it provides an interesting description of the Aharonov-Bohm effect.
\end{abstract}

														
														%
                             
\maketitle


\section{Introduction}\label{sec:int}

One of the main features of the field concept in classical electromagnetism is that it avoids action at a distance. If we have two charged particles, for instance, we can say that the charge of each particle modifies the electromagnetic field configuration around it and that this field acts locally in the charge of the other particle causing a force. But with the prediction \cite{aharonov59} and observation  \cite{chambers60,matteucci85,webb85,tonomura86,peshkin,vanou98,bachtold99,peng10} of the Aharonov-Bohm (AB) effect, in which the interference pattern of charged quantum particles is affected by the electromagnetic field configuration even if they do not interact locally with these fields, the local interaction between charges and fields lost its generality in the description of electromagnetic phenomena. 

Here we go deeper into this subject by reintroducing an alternative expression for the electromagnetic Lagrangian that governs the interaction between a charged particle and applied fields, as proposed by Livens about one century ago \cite{livens16,livens}. This alternative Lagrangian is written in terms of the local superposition of the particle fields with the applied electromagnetic fields, not in terms of the particle charge and of the electromagnetic potentials as is usual \cite{jackson,landau}. We show that the alternative Lagrangian is equivalent to the traditional one in their domain of validity and that it provides an interesting interpretation of the AB effect. We also show that the total Lagrangian for a set of charged particles assumes a simple elegant form with the alternative Lagrangian, what gives an aesthetic support for it. In this way, we believe that it brings new insights for our understanding of the electromagnetic theory.

The paper is organized as follows: In Sec. \ref{sec:lagrang} we describe the alternative expression for the interaction electromagnetic Lagrangian,  in Sec. \ref{sec:tot} we show how the Lagrangian for a set of charged particles assumes a simple form in view of the alternative Lagrangian, in Sec. \ref{sec:ab} we discuss the AB effect in view of this Lagrangian, and in Sec. \ref{sec:conclusion} we present our concluding remarks.

\section{Alternative electromagnetic Lagrangian}\label{sec:lagrang}

The Lagrangian density for the interaction of a charged particle with an applied electromagnetic field can be written as \cite{jackson,landau}
\begin{equation}\label{lagrang}
	\mathcal{L}_{int}=-J_\alpha A^\alpha=-\rho\Phi+\mb{J}\cdot\mb{A},
\end{equation}
where summation over repeated indices is assumed, $J=(c\rho,\mb{J})$ is the current density 4-vector, $A=(\Phi/c,\mb{A})$ is the 4-vector potential corresponding to the applied fields and $c$ is the speed of light in vacuum. For a particle with charge $q$ and velocity $\mb{v}$ at the position $\mb{r}_0$, we have $\rho=q\delta(\mb{x}-\mb{r}_0)$ and $\mb{J}=q\mb{v}\delta(\mb{x}-\mb{r}_0)$. The interaction Lagrangian is given by the volume integral of the above Lagrangian density. The Lorentz force law is deduced when the Euler-Lagrange equations are used in the total Lagrangian, which is the sum of the interaction and free particle Lagrangians \cite{jackson,landau}.

In this work we want to check if the Lagrangian density of Eq. (\ref{lagrang}) is equivalent to a Lagrangian density which is written in terms of the local superposition of the electric and magnetic fields generated by a charged particle, $\mb{E}^{(p)}$ and $\mb{B}^{(p)}$, and the applied electric and magnetic fields, $\mb{E}^{(0)}$ and $\mb{B}^{(0)}$. Calling $F^{(p)}$ the electromagnetic tensor of the fields generated by the particle and $F^{(0)}$ the electromagnetic tensor of the applied fields, Livens proposed the following alternative interaction Lagrangian density \cite{livens16,livens}:
\begin{equation}\label{L'}
\mathcal{L}_{int}'=\frac{1}{2\mu_0}F^{(p)}_{\alpha\beta}F^{(0)\alpha\beta}=-\varepsilon_0\mb{E}^{(p)}\cdot\mb{E}^{(0)}+\frac{1}{\mu_0}\mb{B}^{(p)}\cdot\mb{B}^{(0)},
\end{equation}
where  $\varepsilon_0$ and $\mu_0$ are the electric permittivity and magnetic permeability of free space respectively, with $c=1/\sqrt{\mu_0\varepsilon_0}$. If we write $\mb{E}^{(0)}=-\bs{\nabla}\Phi-\partial\mb{A}/\partial t$ and $\mb{B}^{(0)}=\bs{\nabla}\times\mb{A}$, we can write the action term resulting from the alternative interaction Lagrangian as
\begin{equation}\label{A'}
	\int dt \int d^3x \mathcal{L}_{int}' =S'_1+S'_2
\end{equation}
with
\begin{equation}\label{A1}
		S'_1=\int dt \int d^3x \;\varepsilon_0\mb{E}^{(p)}\cdot\left(\bs{\nabla}\Phi+\frac{\partial\mb{A}}{\partial t}\right)
\end{equation}
and
\begin{equation}\label{A2}
	S'_2=\int dt \int d^3x \frac{1}{\mu_0}\mb{B}^{(p)}\cdot(\bs{\nabla}\times\mb{A}).
\end{equation}
The volume integrals above must be performed through the whole space where the fields are nonzero. Integrating by parts the volume integral of the first term of Eq. (\ref{A1}) and using Gauss' law $\bs{\nabla}\cdot\mb{E}^{(p)}=\rho/\varepsilon_0$, we obtain
\begin{equation}\label{A1b}
		S'_1=\int dt \int d^3x \left[-\rho \Phi +\varepsilon_0\mb{E}^{(p)}\cdot\frac{\partial\mb{A}}{\partial t}\right].
\end{equation}
Using the vector identity $\mb{B}^{(p)}\cdot(\bs{\nabla}\times\mb{A})=\mb{A}\cdot(\bs{\nabla}\times\mb{B}^{(p)})+\bs{\nabla}\cdot(\mb{A}\times\mb{B}^{(p)})$ and the Maxwell-Amp\'ere law $\bs{\nabla}\times\mb{B}^{(p)}=\mu_0\mb{J}+\mu_0\varepsilon_0\partial\mb{E}^{(p)}/\partial t$ in Eq. (\ref{A2}), we obtain
\begin{equation}\label{A2b}
		S'_2=\int dt \int d^3x \left[\mb{J}\cdot\mb{A} +\varepsilon_0\mb{A}\cdot\frac{\partial\mb{E}^{(p)}}{\partial t}\right].
\end{equation}
By comparing Eqs. (\ref{A'}), (\ref{A1b}) and (\ref{A2b}) with Eq. (\ref{lagrang}) we can write
\begin{equation}\label{A'2}
	\int d^4x \mathcal{L}_{int}' =\int d^4x \mathcal{L}_{int} +\int d^4x\;\varepsilon_0 \frac{\partial}{\partial t} [\mb{A}\cdot\mb{E}^{(p)}],
\end{equation}
with $d^4x\equiv d^3x dt$. In the nonrelativistic limit, we have $\mb{E}^{(p)}=\bs{\nabla}[q/(4\pi\varepsilon_0|\mb{x}-\mb{r}_0|)]$, where $q$ is the particle charge and $\mb{r}_0$ its position. Substituting on the second term on the right side of Eq. (\ref{A'2}), integrating by parts the volume integral and using the Coulomb gauge $\bs{\nabla}\cdot\mb{A}=0$, we see that this term is zero.  In the following we show that this term is also zero in the relativistic case in the domain of validity of the Lagrangian density from Eq. (\ref{lagrang}).

The complete treatment of a charged particle interacting with the electromagnetic field is extremely complex \cite{jackson,landau,rohrlich}. This is because the moving particle changes the electromagnetic field configuration, what changes the particle equation of motion in a recursive way. The Lagrangian density of Eq. (\ref{lagrang}) does not take into account the back action of the electromagnetic fields generated by the particle on its own equation of motion. So this Lagrangian density is valid as long as we can neglect the terms of the field generated by the particle that depend on the particle acceleration and on higher order time derivatives of its position, since these terms result in back action \cite{jackson,landau,rohrlich} (the radiation reaction). In this work we investigate if the Lagrangian density of Eq. (\ref{L'}) is equivalent to the Lagrangian density of Eq. (\ref{lagrang}) to the extent that the Lagrangian density of Eq. (\ref{lagrang}) is valid. So we can neglect the terms from $\mb{E}^{(p)}$ in Eq. (\ref{A'2}) that depend on the acceleration and on higher order time derivatives of the particle position, since we will disregard their influence on the particle motion. So we will consider that $\mb{E}^{(p)}$ is the field generated by a particle moving with constant speed, depending only on the particle velocity $\mb{v}$ and on its position $\mb{r}_0$. 


The term on the left and the first term on the right side of Eq. (\ref{A'2}) are Lorentz scalars, since the Lagrangian densities of Eqs. (\ref{lagrang}) and (\ref{L'}) are manifestly Lorentz scalars and the four-dimensional volume element $d^4x$ is invariant under Lorentz transformations. So the second term on the right side of Eq. (\ref{A'2}) must also be a Lorentz scalar, such that its value must be the same in any inertial reference frame. Since $\mb{A}\cdot\mb{E}^{(p)}$ is a function of the independent variables $\mb{x}\equiv(x,y,z)$ and $t$, the partial derivative on time in Eq. (\ref{A'2}) can be substituted by the total derivative. If we compute the time integral between times $t_1$ and $t_2$ in a reference frame $\mathcal{S}$ where the particle velocity at time $t_1$ is zero, we have
\begin{eqnarray}\nonumber
&&	\int_{t_1}^{t_2} dt\int d^3x\;\varepsilon_0 \frac{d}{d t} \left[\mb{A}(\mb{x},t)\cdot\mb{E}^{(p)}(\mb{x},t)\right]=\\&&=\int d^3x\;\varepsilon_0\mb{A}(\mb{x},t_2)\cdot\mb{E}^{(p)}(\mb{x},t_2),
\end{eqnarray}
since, as discussed before, the volume integral of $\mb{A}\cdot\mb{E}^{(p)}$ is zero in the nonrelativistic regime corresponding to $t_1$. Now consider another reference frame $\mathcal{S}'$ in which the particle velocity at $t_2'$ is zero. Since the action must have the same value in both frames, we have 
\begin{eqnarray}\label{A'3-0}\nonumber
&&\int d^3x\;\varepsilon_0\mb{A}(\mb{x},t_2)\cdot\mb{E}^{(p)}(\mb{x},t_2)=\\  &&=-\int d^3x'\;\varepsilon_0\mb{A}'(\mb{x}',t_1')\cdot\mb{E}^{(p)'}(\mb{x}',t_1'),
\end{eqnarray}
where the primed variables in $\mathcal{S}'$ are obtained from the unprimed variables in $\mathcal{S}$ through the correspondent Lorentz transformation.
The above equation must be valid for any $\mb{A}'(\mb{x}',t')$. In particular, it must be valid when $\mb{A}'(\mb{x}',t_1')=0$, such that both sides of Eq.  (\ref{A'3-0}) must always be zero, implying that second term on the right side of Eq. (\ref{A'2}) must be zero under the above considerations.  So the Lagrangian density of Eq. (\ref{L'}) is equivalent to the Lagrangian density of Eq. (\ref{lagrang}) under the domain of validity of this Lagrangian density. 

\section{Total Lagrangian for a set of charged particles}\label{sec:tot}

Now we show how the Livens' interaction electromagnetic Lagrangian density from Eq. (\ref{L'}) results in an elegant expression for the total Lagrangian for a set of charged particles that interact among themselves. Let us first describe the Lagrangian of a single free particle. The mass of each charged particle has an electromagnetic contribution that results from the energy present in the electromagnetic field it generates \cite{rohrlich,feynman}. We will call the electromagnetic contribution for the mass $m_e$, with the total mass being $m=m_e+m'$ and $m'$ resulting from the contribution of extra fields of unknown origin. These extra fields are necessary for the particle stability in any classical model for a charged particle \cite{rohrlich,feynman}. The reason is simple: if the particle was purely electromagnetic, the electric charge that constitutes it would spread out due to the electrostatic repulsion. To keep the charge within the particle, there must be some kind of cohesive forces, that were first postulated by Poincar\'e \cite{rohrlich}. We will assume the existence of fields associated with these cohesive forces,  that from now on will be referred to as Poincar\'e fields, and their energy together with the energy of their interaction with the electromagnetic field are responsible to the contribution $m'$ to the particle mass. Note that $m$ and $m_e$ are positive, but $m'$ can be negative. So in the particle rest frame we can write:
\begin{equation}
	mc^2=m_ec^2+m'c^2=\int d^3r_0 \frac{\varepsilon_0}{2}E_0^2+\int d^3r_0 U',
\end{equation}
where $\mb{E}_0$ represents the electric field generated by the particle at rest (with $E_0\equiv|\mb{E}_0|$), $U'$ is the energy density associated to the Poincar\'e  fields and their interaction with the particle electromagnetic field, and $d^3r_0$ represents the volume element in the particle rest frame.

The relativistic Lagrangian for a free particle  with velocity $\mb{v}$, given by $-mc^2\sqrt{1-v^2/c^2}$ \cite{jackson,landau}, can thus be written as 
\begin{equation}\label{free}
	-mc^2\sqrt{1-v^2/c^2}=\int d^3r \frac{1}{4\mu_0}F_{\alpha\beta}F^{\alpha\beta}-\int d^3r U',
\end{equation}
where $F$ represents the electromagnetic tensor of the particle fields, as long as $U'$ be a Lorentz scalar. This is easy to understand if we note that the values of the Lorentz scalars $F_{\alpha\beta}F^{\alpha\beta}$ and $U'$ are the same in any inertial reference frame, but the volume element $d^3r$ suffers Lorentz contraction such that we have $d^3r=\sqrt{1-v^2/c^2}d^3r_0$. 

Let us now consider the total electromagnetic Lagrangian for a set of $N$ charged particles using Eq. (\ref{free}) for the Lagrangian of each free particle and Eq. (\ref{L'}) for the interaction Lagrangian between each particle and the electromagnetic field generated by the other particles. Calling $F^{(i)}$ the electromagnetic tensor of the fields of particle $i$, $F^{(0i)}\equiv\sum_{j\neq i}F^{(j)}$ the electromagnetic tensor of the fields of all particles with the exception of particle $i$ and $-U'^{(i)}$ the Lagrangian density associated to the Poincar\'e fields for particle $i$, we have
\begin{eqnarray}\label{tot}\nonumber
	L&=&\sum_i\int d^3r \left[\frac{F^{(i)}_{\alpha\beta}F^{(i)\alpha\beta}}{4\mu_0}-U'^{(i)}+\frac{F^{(i)}_{\alpha\beta}F^{(0i)\alpha\beta}}{2\mu_0}\right]=\\
	&=&\int d^3r\left[\frac{F^{(T)}_{\alpha\beta}F^{(T)\alpha\beta}}{4\mu_0}-U'^{(T)}\right],
\end{eqnarray}
where $F^{(T)}\equiv\sum_iF^{(i)}$ represents the electromagnetic tensor of the total electromagnetic field  and $-U'^{(T)}\equiv-\sum_iU'^{(i)}$ represents the Lagrangian density associated to the total Poincar\'e fields. Note the simplicity of the above Lagrangian. This gives an aesthetic support for the alternative electromagnetic Lagrangian from Eq. (\ref{L'}). 

\section{Aharonov-Bohm effect}\label{sec:ab}

In this section we revise the AB effect and describe it in view of the alternative electromagnetic Lagrangian. In a seminal paper, Aharonov and Bohm showed that a charged quantum particle may be influenced by the existence of electromagnetic fields even if it does not directly interact with these fields \cite{aharonov59}. In Fig. 1 two AB experimental schemes are represented. In Fig. 1(a) we have an electron interferometer with a solenoid enclosed by the two possible paths. The magnetic field $\mb{B}^{(0)}$ is confined inside the solenoid and the vector potential $\mb{A}$, linked to the magnetic field through the relation $\mb{B}^{(0)}=\bs{\nabla}\times\mb{A}$, circulates around the solenoid as shown in the figure. The electrons can propagate only outside the solenoid, such that they experience no magnetic field and no Lorentz force, but are subjected to the vector potential. The presence of the vector potential in this region causes a phase difference between the probability amplitude associated with the two interferometer paths. The electron wavefunction in each path accumulates a phase given by $-S/\hbar$, where $S$ is the action for each path and $\hbar$ is Planck's constant divided by $2\pi$ \cite{aharonov59}. The action is given by the time integral of the total Lagrangian, and by using Eq. (\ref{lagrang}) it can be shown that this phase difference is $q\phi_0/\hbar$, where $\phi_0$ is the magnetic flux in the solenoid \cite{aharonov59}. So the electron interference pattern depends on the enclosing magnetic flux even if the the electron only propagates in regions with zero magnetic field. Experiments with different systems confirmed this theoretical prediction, with electrons propagating in free space \cite{chambers60,tonomura86,peshkin}, metals \cite{webb85,vanou98}, carbon nanotubes \cite{bachtold99}, and nanoribbons \cite{peng10}. In most of the cited experiments the charged particles experience a magnetic field during the propagation. But in the experiments of Tonomura \textit{et al.} \cite{tonomura86} a toroidal magnet covered with a superconductor material was used to reduce the magnetic field in the electrons paths to negligible values and the AB effect is still observed.

In Fig. 1(b) we have the electric version of the AB effect. The experiment should be made with electrons sent one by one to the interferometer. The scalar potentials of the conductor tubes in each path are zero when the electron is outside them. When the electron is in a superposition of being inside each of the tubes, the potential of the tubes is varied and comes back to zero before the electron wavefunction exits each of them. In this case the electrons experience no electric field and no Lorentz force, but experience different electric potentials in each path. Again, this results in a phase difference between the paths. If the potential of the tube of path $a$ is $\Phi_a(t)$ and the one of path $b$  is $\Phi_b(t)$,  by using Eq. (\ref{lagrang}) it can be shown that the phase difference is $\int dt[\Phi_a(t)-\Phi_b(t)]q/\hbar$, affecting the interference pattern \cite{aharonov59}. This behavior was also experimentally verified \cite{matteucci85,vanou98}, although not with the charged particles propagating only in free-field regions.

\begin{figure}\begin{center}
  \includegraphics[width=8.5cm]{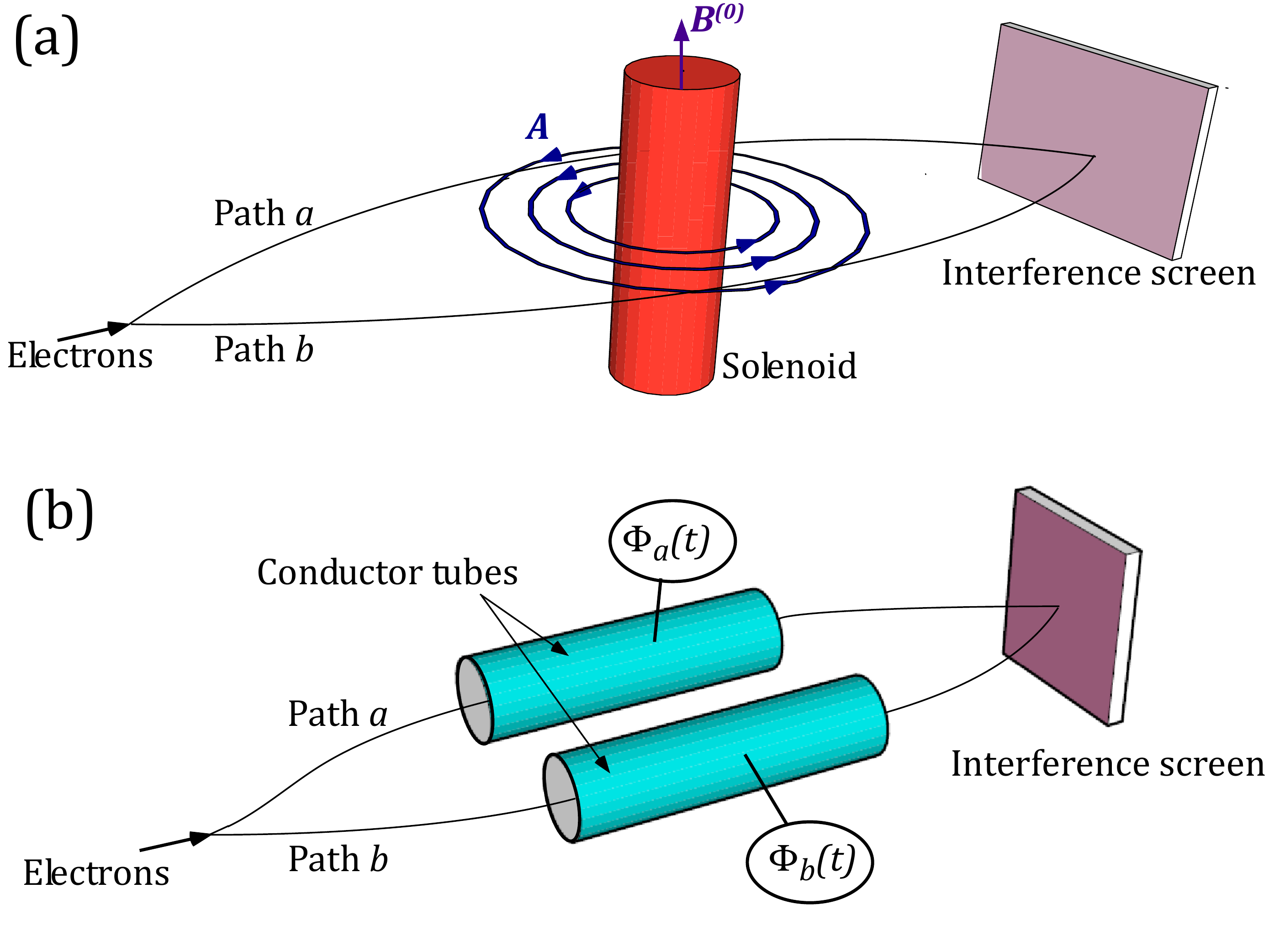}\\
  \caption{Aharonov-Bohm (AB) schemes. (a) Magnetic AB scheme. The interference pattern of an electron beam on a screen depends on the magnetic flux in the solenoid even if the electrons only propagate in regions with no electromagnetic fields. (b) Electric AB scheme. Electrons are sent to the interferometer one by one. The scalar potentials $\Phi_a(t)$ and $\Phi_b(t)$ are different from zero only while the electron wavefunction is in a superposition of being inside each of the tubes. The interference pattern depends on the potentials even if the electrons only propagate in regions with no electromagnetic fields.}
\label{fig1}
 \end{center}\end{figure}

We can now interpret the AB experiments illustrated in Fig. 1 using the alternative Lagrangian density from Eq. (\ref{L'}). In the experiment of Fig. 1(a), we can see that when the electron propagates through path $a$ ($b$), it generates a magnetic field inside the solenoid that has a positive (negative) component in the direction of $\mb{B}^{(0)}$. This results in a positive (negative) contribution for the action term resulting from Eq. (\ref{L'}), generating a phase difference between the paths. In the experiment of Fig. 1(b), considering the potential zero when there is no charge on the tube, when the electron is inside a tube with a positive (negative) potential, such that the tube has a positive (negative) charge, the electric field generated by the electron has always a negative (positive) component in the direction of $\mb{E}^{(0)}$ outside the tube. Using the same reasoning as before we conclude that different potentials on the tubes produce a phase difference between the paths.  

The idea of associating the AB effect to the interaction of the electron electromagnetic field with the solenoid in Fig. 1(a) is not new, see Ref. \cite{peshkin}. Liebowitz \cite{liebowitz65} and Boyer \cite{boyer73,boyer02}, for instance, suggested that a back action force could act on the electron due to the magnetic energy that results from the overlap of the electron and solenoid magnetic fields, predicting a lag between the electron paths that would explain the AB effect. Recent experiments seem to rule out this back action force hypothesis  \cite{caprez07}, but there are still counter arguments  \cite{boyer08}. Peshkin computed the angular momentum that results from the superposition of the electron electric field with the solenoid magnetic field and used the angular momentum quantization in the system as an argument for the existence of the AB effect \cite{peshkin81}. More recently, Vaidman deduced the AB phase using a quantum mechanical treatment for the charges of the solenoid interacting with the electron field \cite{vaidman12}. But our work treats the issue in a more fundamental level than the cited works, by suggesting a modification on the expression of the electromagnetic Lagrangian. Also, the theoretical descriptions of these previous works assume that the electromagnetic fields produced by the electron penetrates inside the solenoid of Fig. 1(a). But in the experiments of Tonomura \textit{et al.} \cite{tonomura86}, where the solenoid  is replaced by a toroidal magnet covered with a superconductor material, the superconductor does not let the electrons field to penetrate inside the toroidal magnet, such that these previous works cannot explain the observation of the AB phase in this experiment. As we discuss in the following, the alternative formulation also predicts the AB phase in this case.

In the experiments of Tonomura \textit{et al.} \cite{tonomura86}, let us call $\mb{B}^{(p)}$ the magnetic field generated by the electron, $\mb{B}^{m}$ the magnet magnetic field and $\mb{B}'$ the magnetic field generated by the charges of the superconductor inside the toroidal magnet. The fact that the superconductor cancel the influence of external fields inside it can be stated, according to the superposition principle, as $\mb{B}'=-\mb{B}^{(p)}$. According to Eq. (\ref{L'}), the alternative Lagrangian involves the scalar product of $\mb{B}^{(p)}$ with the other fields, in this case:  $\mb{B}^{(p)}\cdot\mb{B}^{(0)}=\mb{B}^{(p)}\cdot\mb{B}^{m}+\mb{B}^{(p)}\cdot\mb{B}'=\mb{B}^{(p)}\cdot\mb{B}^{m}-|\mb{B}^{(p)}|^2$. But note that the sign of $-|\mb{B}^{(p)}|^2$ does not depend on which path the electron takes in the scheme of Fig. 1(a). So the term that contributes to the AB phase is $\mb{B}^{(p)}\cdot\mb{B}^{m}$, as if the superconductor wasn't there.

\section{Discussion}\label{sec:conclusion}

In summary, we have reintroduced Livens' alternative expression for the electromagnetic Lagrangian density that governs the interaction of charged particles with applied fields, written in terms of the local superposition of the particle and applied electromagnetic fields. The alternative Lagrangian gives rise to a simple elegant expression for the total Lagrangian of a set of charged particles. This change on the expression of the electromagnetic Lagrangian may lead us to a dramatic change in our understanding of the electromagnetic interactions. Instead of saying that the applied electromagnetic fields interact locally with the particle charge causing a force, we could say that the applied fields interact locally with the particle fields causing a change on the particle motion. The concept of electric charge thus assumes a secondary role, as the particle charge would not be directly present in the description of the fundamental electromagnetic interactions. In this view, electric charges and currents could be seen as auxiliary tools for performing the calculations just like the electromagnetic potentials, since according to Eq. (\ref{L'}) the fundamental interactions would be governed by the electromagnetic fields directly. In this sense, the electric charge of a particle would effectively represent a stable electromagnetic field configuration that moves in space. This view also naturally incorporates the Aharonov-Bohm effect, which must be seen either as a nonlocal interaction between the particle charge and the electromagnetic fields or as a local interaction between the particle charge and the electromagnetic potentials in the traditional interpretation of electromagnetism. In the alternative view, the Aharonov-Bohm effect is also the result of the local superposition of the particle and applied electromagnetic fields. 

\section*{Acknowledgements}
I would like to acknowledge Kirk McDonald for very useful discussions and for calling my attention to the works of Livens \cite{livens16,livens}. I would also like to acknowledge Ricardo Schor for useful discussions and J\'ulia Parreira for useful comments on the manuscript. This work was supported by the Brazilian agencies CNPq and CAPES.


\end{document}